\documentclass{elsart}
 \usepackage{graphicx}
 \usepackage{epsfig}

\usepackage{amssymb}

\def\beq{\begin{equation}}
\def\eeq{\end{equation}}
\def\beqn{\begin{eqnarray}}
\def\eeqn{\end{eqnarray}}

\newcommand{\ee}{e^{+}e^{-}}
\newcommand{\mumu}{\mu^{+}\mu^{-}}

\newcommand{\qts}{\mbox{${\bf q}^2_T$}}
\newcommand{\qt}{\mbox{${\bf q}_T$}}

\newcommand{\SKIP}[1]{}

\newcommand{\CMS}{\mbox{cm$^{-2}\,$s$^{-1}$}}

\begin{document}

\begin{frontmatter}

\title{
\hspace*{9cm} {\large\rm Budker INP 2002-48} \\[1cm]
Precision luminosity measurement at
LHC using two-photon production of $\mu^{+}\mu^{-}$ pairs.
}


\author{A.G. Shamov, V.I. Telnov}
\thanks{Talk at the VIII Intern. Conference on
  Instrumentation for Colliding Beam Physics, Novosibirsk, Russia,
  Feb.~28-March~6, 2002} 
\address{Budker Institute of Nuclear Physics, 630090 Novosibirsk,
  Russia \\ 
A.G.Shamov@inp.nsk.su,V.I.Telnov@inp.nsk.su }

\begin{abstract}
 The application of the two-photon process
 $pp \to pp + \mumu$ for the luminosity measurements at LHC with
 the ATLAS detector is considered. The expected accuracy of
 the absolute offline luminosity determination 
 is 1 $\div$ 2 \% for the luminosity range of  $10^{33} \!\div\! 
 10^{34}$ \CMS. The preliminary cross section estimates done for LHCb promise
the same
level of the luminosity measurement accuracy at $ L = 2 \cdot 10^{32}$ \CMS.
\end{abstract}

\end{frontmatter}
\section{Introduction}
 The possibility to use the two-photon pair production for luminosity
measurements at hadron colliders was first considered in \cite{NPHYS}.
It was shown
that the cross section of the process $pp \to pp + e^{+}e^{-}$ for
the forward region can be
calculated within QED with an accuracy of 1\% or better allowing for
a precise luminosity determination using the number of events observed and
the calculated cross section: $L = N^{obs}/\sigma^{obs}$.

The method suggested in \cite{NPHYS} provides for both the high rate monitoring
and the absolute luminosity measurements but
requires a dedicated apparatus allowing for the 
$e^{+}e^{-}$-pair detection in the pseudorapidity range of $|\eta| >
$ 5 and capable of the $p_T$ measurements with the accuracy of a few
MeV. This approach has never been used.
 The kinematic domain of the central rapidity pairs is
much more attractive from the experimental point of view though
the corresponding cross section is smaller and some additional
theoretic uncertainties appear.

The application of the centrally produced $e^{+}e^{-}$-pairs with
$p_T = 0.4 \div 1 $ GeV 
for
luminosity measurements at LHC with ATLAS was proposed in \cite{VT,TP}.
The detail study of the trigger issues and the background conditions
has shown that using of $pp \to pp + \mumu$ with high-$p_T$
muons looks favorable.

Due to the high $p_T$ threshold of LHC experiments ($\geq$ 1 GeV), the
cross section of $pp \to pp + \mumu$ is small ($1\div 15$ pb)
therefore this process can not be used for the luminosity monitoring
but provides for the absolute calibration of any stable high-rate
monitor with the statistical
accuracy of 1 $\div$ 2 \% after of a few months operation.

Below the proposal for the ATLAS offline luminosity
determination 
using the two-photon production of $\mumu$-pairs
\cite{MASL,TDR,Helsinki}
is reported and
the preliminary signal and background cross section estimates
for the LHCb detector are presented.

It should be mentioned that the application of the forward 
$e^{+}e^{-}$-pairs is under study in ATLAS \cite{KP,TDR,Helsinki}.  
The possibility to employ  $pp \to pp + \mu^{+}\mu^{-}$ for luminosity
measurements was manifested in \cite{COURAU}
without a detail consideration of theoretical uncertainties and  detector
related questions.
\vspace{-5mm}
\section{On possibility of the cross section calculation}
\vspace{-5mm}
A review of the two-photon leptoproduction can be found
elsewhere \cite{PHYSREP}.
Main features of this process can be illustrated
in the Equivalent Photon Approximation (E.P.A.) allowing one to express 
its cross section via the cross section of the pair production
by virtual photons with the equivalent photon spectra
\beq\label{S}
d\sigma = \sigma_{\gamma^{*}\gamma^{*} \to LL} \: dn_1 \: dn_2
\eeq
For colliding leptons one has \cite{PHYSREP}
\beq\label{N}
  dn_{QED} 
    = \frac{\alpha}{\pi} \:
            \frac{d\omega}{\omega} \:
            \frac{dq^2}{q^2} \:
            \left( 
               1 - \frac{q^2_{min}}{q^2}
            \right)
    = \frac{\alpha}{\pi} \:
            \frac{d\omega}{\omega} \:
            \frac{\qts\,d\qts}{\left(\omega^2/\gamma^2+\qts\right)^2} 
            \, 
\eeq
where $q\,(\omega,\bf{q})$ is the four-momentum of the virtual photon
and $\gamma = E/m$ is the Lorentz factor of the colliding particle
(we assume $\omega \ll E$).

The cutoff of the spectrum (\ref{N}) occurs at $|\qt| \sim W$
due to the $q^2$ dependence of $\sigma_{\gamma^{*}\gamma^{*} \to LL}$ 
($W$ is the invariant mass of the pair produced).
The characteristic value of the the total transverse
momentum of the pair produced $p_T \sim (\omega_1 + \omega_2)/\gamma \sim $
10 MeV in the luminosity measurements context.

When $pp$ collide, the proton can dissociate during the photon
emission therefore the elastic and the inelastic two-photon processes
should be distinguished.

\begin{figure}[h]
\begin{center}
    \vspace*{-1.8cm}
    \mbox{
      \hspace*{-1.8cm}
      \mbox{\epsfysize=6.5cm\epsffile{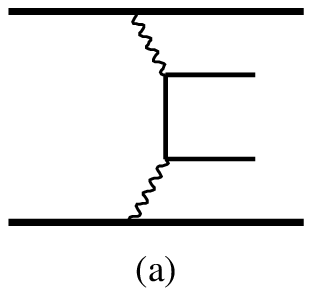}}
      \hspace*{-3.2cm}
      \mbox{\epsfysize=6.5cm\epsffile{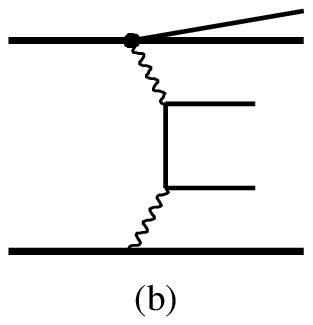}}
      \hspace*{-3.2cm}
      \mbox{\epsfysize=6.5cm\epsffile{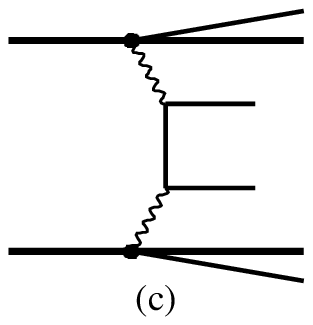}}
      } 
    \vspace*{-2.4cm}
\end{center}
  \caption{Two-photon production of lepton pairs in $pp$ collisions: 
    (a) - elastic, (b,c) - inelastic
    \label{fig:fgs}}
\end{figure}

In the elastic case Fig. (\ref{fig:fgs},a) only the small modification
of the equivalent photon spectrum is required:
 \beq\label{Nel}
dn_{el} = dn_{QED} \times
     \frac{G_E^2 - q^2/4m_p^2 \: G_M^2}{1 - q^2/4m_p^2}
 \eeq
\vspace*{-18pt}
($G_E$, $G_M$ are the electromagnetic form factors).

For the inelastic processes Fig. (\ref{fig:fgs},b), (\ref{fig:fgs},c)
one has for each inelastic vertex \cite{PHYSREP,NPHYS}
\beq\label{Ninel}
  dn_{inel} = dn_{QED} \times
  \frac{W_2(q^2,M^2)}{2m_p} \, dM^2
  \approx  dn_{QED} \times
     \frac{|q^2|}{4\pi^2\alpha} \:\,
        \frac{ \sigma^{\gamma p}_{T}+\sigma^{\gamma p}_{S} }
                       {M^2-m_p^2} \, dM^2
 \eeq
 where $M$ is the invariant mass of the hadronic system,
 $W_2$ is the inelastic scattering 
structure function and
$\sigma^{\gamma p}_{T}$, $\sigma^{\gamma p}_{S}$ are
 the $\gamma p$ cross sections known
from the photo- and electro-production experiments.
Unlike (\ref{N}) and (\ref{Nel}),
 the expression (\ref{Ninel}) is not singular
for $q^2 \to 0$ thus the characteristic pair $p_T$ for the
inelastic production is not small ($\sim$ 250 MeV).

\begin{figure}[h]
\begin{center}
    \vspace*{-1.8cm}
    \mbox{
      \hspace*{-1.8cm}
      \mbox{\epsfysize=6.5cm\epsffile{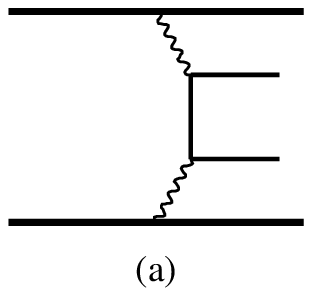}}
      \hspace*{-3.2cm}
      \mbox{\epsfysize=6.5cm\epsffile{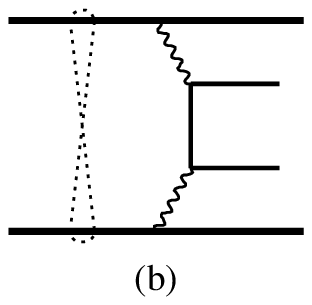}}
      \hspace*{-3.2cm}
      \mbox{\epsfysize=6.5cm\epsffile{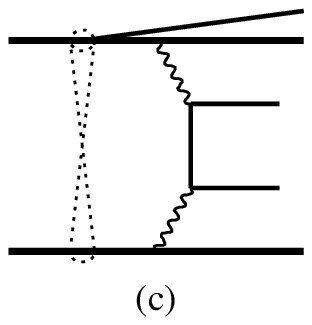}}
      } 
    \vspace*{-2.4cm}
\end{center}
  \caption{Modification of the elastic two-photon process by
rescattering (final state rescattering is not shown).
    \label{fig:rescat}}
\end{figure}

Diagrams of Fig. \ref{fig:fgs} reflect the strong interaction
inside the single
proton, besides, the 
strong interaction between colliding protons, so called {\it rescattering},
should be taken into account (Fig. \ref{fig:rescat}).

In the ATLAS luminosity measurement context,
the rescattering has been considered in \cite{KHOZE}.
It was shown that the elastic strong interaction Fig. (\ref{fig:rescat},b)
just modifies the phase of the matrix element and does not change
the cross section. The inelastic strong interaction Fig.
(\ref{fig:rescat},c)
reduces the yield of $pp \to pp + \mumu$ on behalf of
$pp \to X + \mumu$.
In the experimental setup when only the muon pair is detected,
the total dimuon cross section remains unchanged.
To reduce the dimuon background from Drell-Yan process and
the hadron decays, the absence of extra particles in the dimuon
vertex is necessary. In this case

\beq
 \sigma_{\gamma\gamma}^{pp \to X + \mu\mu} =
      \sigma_{\gamma\gamma}^{el}
        + \epsilon_{inel} \cdot \sigma_{\gamma\gamma}^{inel}
          - (1-\epsilon_{rescat}) \cdot \sigma_{\gamma\gamma}^{rescat}
 \eeq
where
$\epsilon_{inel}, \epsilon_{rescat}$ are the probabilities that the
event passes the vertex cut; for LHCb and ATLAS
$\epsilon_{inel}, \epsilon_{rescat} > 0.8$.

The first two terms can be estimated using the E.P.A. formulae
\beq\label{Sel}
    d\sigma_{\gamma\gamma}^{el}  =  dn_{el,1} \: dn_{el,2}
     \: \cdot \sigma_{\gamma\gamma \to \mumu} \nonumber 
\eeq
\beq\label{Sinel}
    d\sigma_{\gamma\gamma}^{inel} =  \left( 
          dn_{el,1} \: dn_{inel,2} +
          dn_{inel,1} \: dn_{el,2} +
          dn_{inel,1} \: dn_{inel,2} 
     \right) \cdot
      \sigma_{\gamma\gamma \to \mumu} \nonumber
\eeq
$\sigma_{\gamma\gamma}^{inel} \sim \sigma_{\gamma\gamma}^{el}\:$
for the pair mass $W > 1$ GeV
in absence of cuts on the total pair $p_T$.

The $\sigma_{\gamma\gamma}^{rescat}$ can not be expressed in terms of
the equivalent photon spectra. Numerically,
 $\,\sigma_{\gamma\gamma}^{rescat} \sim 0.1\%\:\, $ of
$\:\sigma_{\gamma\gamma}^{el}\:$
 for $p_T$(pair) $<$ 30 MeV \cite{KHOZE}.

 The M.C. calculations performed by us
for the ATLAS conditions has demonstrated that 
$\,\:\sigma_{\gamma\gamma}^{rescat} \sim 0.1 
\,\sigma_{\gamma\gamma}^{inel}\:\,$
and its dependence of 
on the pair $p_T$ is similar to that of
$\:\sigma_{\gamma\gamma}^{inel}$. $\,$ So, it causes no additional
problems for the luminosity measurements.
 Below we include the rescattering 
effect in the inelastic cross section.

It should be noted that the rescattering for the inelastic diagrams
(\ref{fig:fgs},b), (\ref{fig:fgs},c) is not considered in \cite{KHOZE}
and is ignored in our cross section calculations. This and the
bad knowledges of
the soft inelastic scattering structure functions make
the inelastic cross section estimates rather uncertain.

\begin{figure}[hb]
\begin{center}
 \vspace*{-10pt}
 \hspace*{-10pt}
 \mbox{\epsfxsize=12.cm\epsffile{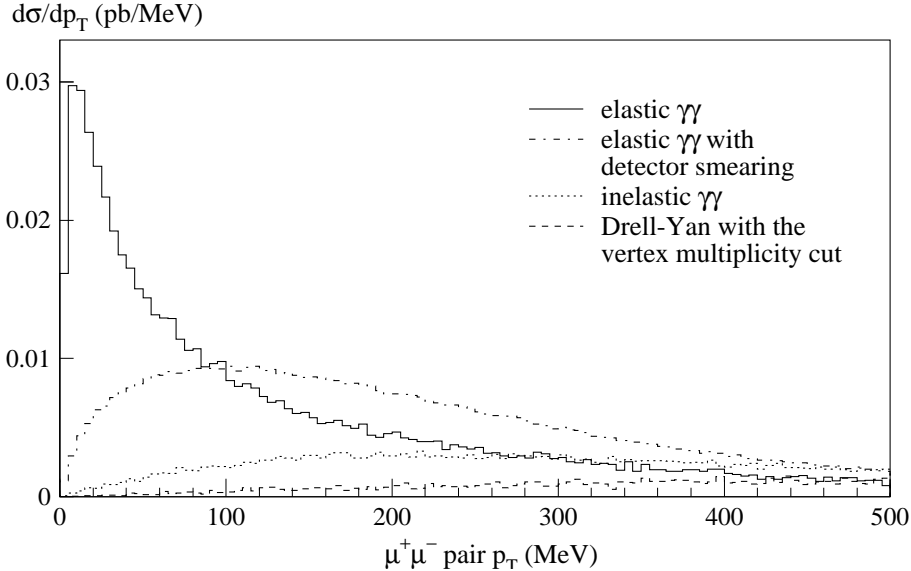}}
 \vspace*{-10pt}
\end{center}
\caption{The differential dimuon cross section $d\sigma/dp_T$
for $pp \to pp + \mumu$ 
as a function of the pair transverse momentum
$p_T$ (ATLAS, $W>$ 12 GeV, $|\eta| <$ 2.3).
\label{fig:pt}}
\end{figure}

For the forward $\ee$-pair production the smallness of  the pair mass $W$ 
ensures the smallness of $q^2$ and makes the inelastic contribution
negligible as was stated in \cite{NPHYS}.
The situation is not so simple concerning the muon pair production with
$p_T(\mu) >$ 6 GeV and $W >$ 12 GeV. 
In this case the inelastic
background is not negligible even with severe experimental cuts (see Fig.
\ref{fig:pt}).
However, the distributions of the pair $p_T$ and related variables
are very different allowing one to reduce the background to an acceptable
level and then subtract it as described below. 

The uncertainty in the
residual inelastic contribution is about 30\%, $\:\sim$ 3\% of the
visible signal, does not affect the luminosity
measurement due to the background subtraction. 
The accuracy of the elastic signal calculation $\leq$ 0.5\% is limited by the
accuracy of the proton form factors only. 

The formulae (\ref{Sel}), (\ref{Sinel}) are presented above just for the
illustration. The actual calculations were performed using the event generator
\cite{LPAIR} implementing the exact lowest order matrix element.
The muon bremsstrahlung, affecting 
the essential parameter distributions, was
simulated in the soft photon approximation; other types of radiative
corrections were not applied. It seems sufficient for the current
stage
of the work.
\vspace{-5mm}
\section{Procedure for signal extraction}
\vspace{-5mm}
 The following criteria are proposed for the ATLAS event selection:
\vspace{-4pt}
\begin{enumerate}
\item Two muon tracks of opposite charges (measured in both the
Inner Detector and the muon spectrometer and triggered by the Muon
systems) with  $p_t > 6 \:$ GeV and $|\eta| < 2.2$;
\item Muon pair invariant mass $< 60$  GeV (against $Z^0$);
\item $p_T$ of the muons are equal within 2.5$\sigma$ of the measurement
uncertainty ($\sigma(p_T)/p_T \simeq$ 1.5\% for $p_T < $ 20 GeV);
\item Acollinearity angle $\Theta > 1^{\circ}$ so the muons are
not exactly back-to-back (against the cosmic ray background);
\item Probability of $\chi^2 > 1\%$ for the muon vertex fit;
\item No other charged tracks with $p_t > 0.5 $ GeV, $|\eta| < 2.5$,
   $\sigma(z_v) < 1 $ mm  and a good $\chi^2$ from the muon vertex
($\sigma(z_v)$ is the estimated error in $z$ for the nearest point
to the beam). 
\end{enumerate} 
\vspace{-3pt}
The condition (6) strongly reduces the background from Drell Yan process
and hadron decays but makes the detection 
efficiency dependent 
on the event pile-up probability.
For the longitudinal LHC bunch size of 7.5 cm 
and the luminosity $L = 10^{33} \:$ \CMS,
the loss of two-photon events is about 3\%.

\begin{figure}[ht]
\begin{center}
 \vspace*{-5pt}
 \hspace*{-10pt}
 \mbox{\epsfxsize=11.5cm\epsffile{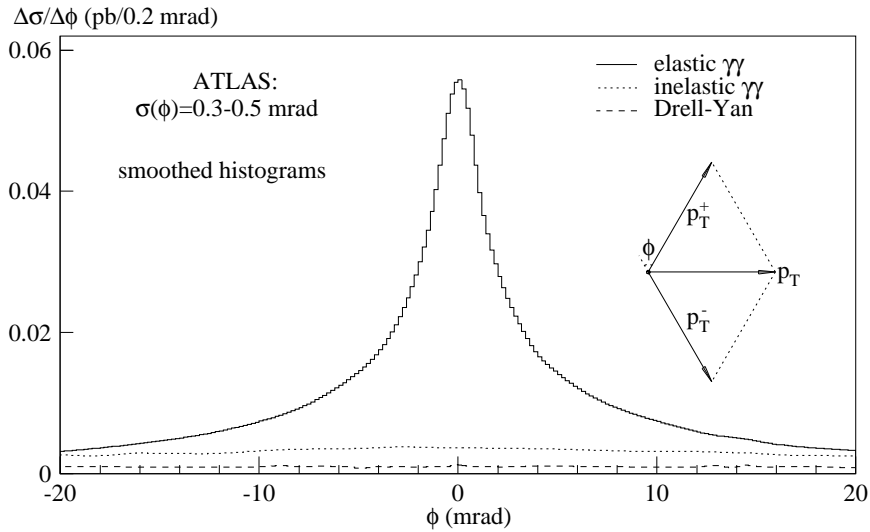}}
 \vspace*{-15pt}
\end{center}
\caption{The acoplanarity distributions for the signal and 
dominant background processes with all cuts applied.
\label{fig:phi}}
\end{figure}

The expected acoplanarity distribution for selected events in shown in Fig.
\ref{fig:phi}. 
(the acoplanarity angle $\phi$ is defined as the angle
between the muon production planes).
The simulation was performed using the particle level
Monte Carlo code \cite{MASL} with
the detector properties parameterized according to the ATLAS specification.
The signal and background cross section estimates 
for the reference region $|\phi| < $ 5 mrad
containing about 50\% of the signal are presented in Table \ref{tab:MM}.

The acoplanarity distribution has the sharp signal peak at $\phi = 0$
and is almost uniform for background processes in the range of
$|\phi| <$ 50 mrad.
Unlike the 
distribution of the pair transverse momentum, it
is not much affected by the detector resolution.
Using this, the final signal extraction can be performed by
fitting the $\phi$ distribution in the range
of $\pm20\!\div\!40$ mrad with the signal shape obtained by Monte Carlo
and the symmetric parabolic background $A-B\phi^2$.
\begin{table}[h]
\caption{ATLAS cross section estimates for the kinematic cuts (1-4) and all
cuts (the expected detection efficiency $\geq$
0.65 is not taken into account).
\label{tab:MM}
}
\vspace{0.2cm}
\begin{center}
\footnotesize
\begin{tabular}{|c|c|c|}
\hline
process    & $\sigma_{kin},\: pb$ & $\sigma_{all}, \: pb$ \\
\hline
{\it signal}     & 1.33 & 1.30$^{\,a}$ \\  
\hline
$\gamma\gamma$(inelastic)   & 0.13 & 0.12$^{\,b}$  \\
Drell-Yan process     & \hspace*{-0.7em} 4.0   & 0.04  \\
Heavy quark decays    & \hspace*{-1.8em} 10.   & 0.01  \\
$\pi/K$ decays        & \hspace*{-0.3em}1.8   & \hspace*{-0.7em} $<$0.001 \\
\hline
{\it background} & \hspace*{-0.9em}15.9 & 0.17 (13\%) \\
\hline
\end{tabular}
\end{center}
{\footnotesize
\vspace*{4pt}
\hspace*{9em} $a$) for $L = 10^{33}\: $\CMS \\*[-2pt]
\hspace*{9em} $b$) including rescattering effect

\vspace{3mm}
}
\end{table}

With the proper modification of the event selection criteria
and the background subtraction $\phi$ interval,
the procedure describe above can be employed in CMS and LHCb
experiments.
\section{On accuracy of ATLAS luminosity determination}
\vspace{-5mm}
The expected statistical accuracy of the luminosity determination
$\approx$ 1.5\% for the integrated luminosity of 10 fb$^{-1}$ (about
4 months of operation at $L=10^{33}\:$ \CMS) 
and the detection (trigger and reconstruction) efficiency of 0.65.

The systematic errors due to the
uncertainty in the proton form factors and the
background subtraction procedure are expected to be $\leq$ 0.5\%.
The pile-up correction can be done using the experimental data.
The systematic error should not exceed 1\% at
$L = 10^{34}$ and is negligible at $L = 10^{33} \: $\CMS.

\begin{figure}[ht]
\begin{center}
 \vspace*{-85pt}
 \mbox{\epsfxsize=11cm\epsffile{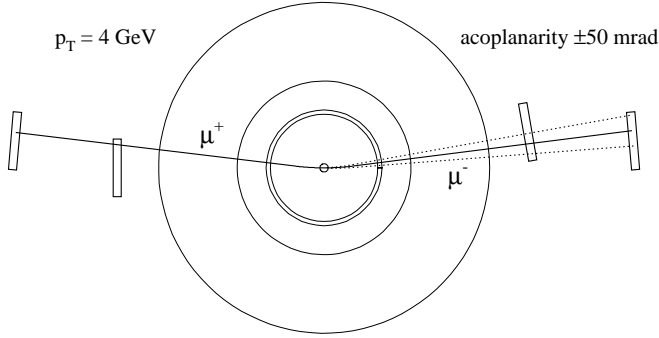}}
 \vspace*{-90pt}
\end{center}
\caption{The common $\mumu$ track in $(\rho,\varphi)$-projection
(the multiple scattering is not taken into account).
\label{fig:trig}}
\end{figure}

The uncertainty of the detection efficiency seems dominating source
of the systematic error. This efficiency can be determined experimentally
using the $pp \to pp + \mumu $ event sample with the accuracy better
than 1\% (it can be keep clean enough with some selection criteria
relaxed).
For the low
luminosity operation ($L=10^{33}\:$ \CMS) the first level
trigger on the single muon with $p_T > $ 6 GeV is foreseen
\cite{ATRG},
and the trigger efficiency can be determined using
the second muon (a minor modification of the second level trigger
software is required).
For the operation at $L=10^{34}\:$ \CMS\ the single muon trigger
with $p_T >$ 6 GeV will have too high rate therefore two such muons
will be required in the first level trigger. The trigger efficiency
for this case is supposed to be known from the low luminosity run.
The other option is triggering
on the common dimuon track in ($\rho,\varphi$)-projection as shown
schematically in Fig. \ref{fig:trig}. This allows 
additionally to reduce the $p_T$ threshold and substantially
increase the signal rate. This needs some extending of the ATLAS
first level trigger.

Without the experimentally measurement efficiency
one has to rely on the Monte Carlo
calculations involving tracking of low momentum
muons in the detector material and in the inhomogeneous magnetic field of
the air core toroid.
In this case the 
accuracy of the luminosity measurement with $pp \to pp + \mumu$ would
hardly be better than 5\%.

\vspace*{-5mm}
\section{Cross section estimates for LHCb}
\vspace*{-5mm}
 To study the potential of the method suggested for the LHCb
experiment, the cross section estimates has been obtained using
 the particle level Monte Carlo code. The particles
were tracked through some model of the LHCb detector without interaction with
the  material. The resolution in $\phi$ and $\theta$ angles was
fixed 
by the number and the width of the strips hitted in the vertex detector.
The total momentum resolution of $0.3 \, p$ \% was assumed \cite{LHCb}.

The following event selection criteria were applied:
\vspace{-4pt}
\begin{enumerate}
\item Two muon tracks of opposite charges 
with  $p_T > 1 \:$ GeV, $E > 8\:$ GeV are detected;
\item Muon pair invariant mass $< 30$  GeV;
\item Difference in the measured values of muon $p_T$ is small: \\
  $|p_T^{+}-p_T^{-}|/(p_T^{+}+p_T^{-}) < 0.13$.
\item No other charged tracks are observed in the event.
\end{enumerate} 

The preliminary results obtained for the reference region
$ |\phi | < 25$ mrad
are presented in Table \ref{tab:LHCB}. 
The observed multiplicity for all background sources except
the two-photon one are $30 \div 60$.
No events passing the condition (4) were generated,
the suppression factors were roughly
estimated using the fits of the multiplicity distributions
for the relaxed cut (3).

\begin{table}[h]
\caption{LHCb cross section estimates for the kinematic cuts (1-3) and all
cuts (the detection efficiency is not taken into account).
\label{tab:LHCB}
}
\vspace{0.2cm}
\begin{center}
\begin{tabular}{|c|r|c|}
\hline
process    & $\sigma_{kin},\: pb$ & $\sigma_{all}, \: pb$ \\
\hline
{\it signal}     & 14.6 & $\approx$14.6 \\ 
\hline
$\gamma\gamma$(inelastic)   & \hspace*{0.2em} 0.6 & \hspace*{0.85em} 0.5  \\
Drell-Yan process     & 93   & \hspace*{0.6em} $<$0.02  \\
  Heavy quarks decays    & 690     & \hspace*{0.6em}    $<$0.02 \\
  $\pi, K$ decays       &  $\sim$50000    & \hspace*{0.3em}  $<$0.3 \\
\hline
{\it background} & 93.6 & $ 0.5 \div 0.84$ ($<$6\%) \\
\hline
\end{tabular}
\end{center}
\end{table}

Due to lower $p_T$ threshold, the estimated LHCb signal cross section 
is more than 10 times larger then that for ATLAS.
The statistical
accuracy of  1\% seems
reachable at the designed luminosity
of $2.10^{32}$ \CMS.

By the same reason the variation of the background level in the
background
subtraction interval $|\phi| < 100 $ mrad is bigger than for ATLAS.
Some optimization of cuts might be necessary to
keep the systematic error small.

\vspace*{-2mm}
\section{Conclusion}
\vspace*{-5mm}
  The method described allows for the ATLAS offline luminosity determination
using the two-photon process $pp \to pp + \mumu$ 
with the accuracy  $\leq$ 2\% 
in the luminosity range of $10^{33} \!\div\! 10^{34} \:$ \CMS.
The preliminary cross section estimates done for LHCb promise
the same
level of the luminosity measurement accuracy at
$ L =  2 \cdot 10^{32}\:$ \CMS.

\vspace*{-5mm}
\section*{Acknowledgments}
\vspace*{-5mm}
The authors would like to thank
I.F. Ginzburg, N. Ellis, D. Froidevaux, P. Jenni, V. Khoze, 
K. Piotrzkowski, M.G. Ryskin, V.G. Serbo
and Yu.A. Tikhonov for support of the work and useful
discussions.

\end{document}